\documentclass{article}

\usepackage{PRIMEarxiv}
\usepackage{amsmath}
\usepackage[utf8]{inputenc} 
\usepackage[T1]{fontenc}    
\usepackage{hyperref}       
\usepackage{url}            
\usepackage{booktabs}       
\usepackage{amsfonts}       
\usepackage{nicefrac}       
\usepackage{microtype}      
\usepackage{lipsum}
\usepackage{fancyhdr}       
\usepackage{graphicx}       
\graphicspath{{media/}}     

\title{Evaluation Metric for Quality Control and Generative Models in Histopathology Images}

\author{
  Pranav Jeevan\thanks{Equal Contribution.} \hspace{10mm} Neeraj Nixon\footnotemark[1]
  \hspace{10mm} Abhijeet Patil 
  \hspace{10mm} Amit Sethi\\
  Department of Electrical Engineering \\
  Indian Institute of Technology Bombay \\
  Mumbai, India\\
  \texttt{\{pjeevan, 20d070056, abhijeetptl, asethi\}@iitb.ac.in} \\
}

\begin{document}
\maketitle

\begin{abstract}
Our study introduces ResNet-L2 (RL2), a novel metric for evaluating generative models and image quality in histopathology, addressing limitations of traditional metrics, such as Fréchet inception distance (FID), when the data is scarce. RL2 leverages ResNet features with a normalizing flow to calculate RMSE distance in the latent space, providing reliable assessments across diverse histopathology datasets. We evaluated the performance of RL2 on degradation types, such as blur, Gaussian noise, salt-and-pepper noise, and rectangular patches, as well as diffusion processes. RL2's monotonic response to increasing degradation makes it well-suited for models that assess image quality, proving a valuable advancement for evaluating image generation techniques in histopathology. It can also be used to discard low-quality patches while sampling from a whole slide image. It is also significantly lighter and faster compared to traditional metrics and requires fewer images to give stable metric value.
\end{abstract}

\section{Introduction and Background}
\label{sec:intro}

\begin{figure}[htb]
  \centering
   \centerline{\includegraphics[scale=0.8]{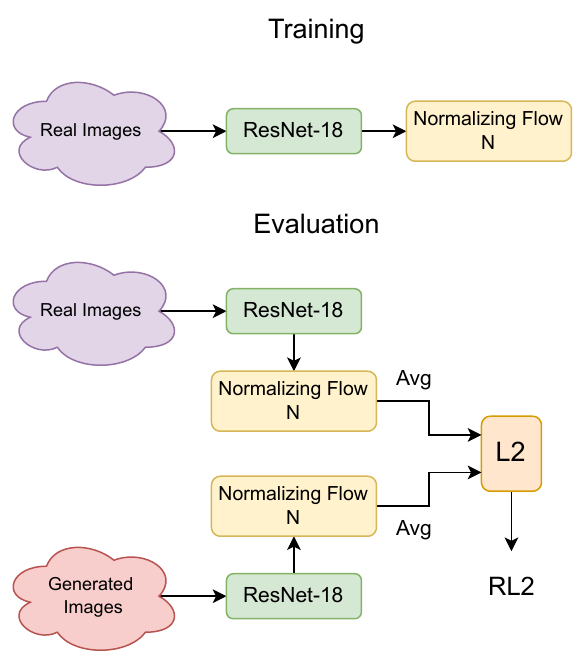}}
\caption{The process of computing RL2. In the training phase, the ResNet-normalizing flow network is trained on the given real (high-quality) images. In the evaluation phase, real and generated (or evaluation) images are passed through the network, and the L2 distance between mean of latent vectors of real and generated (or evaluation) images is used as the final metric.}
\label{fig:rl2}
\end{figure}

Generative models, particularly generative adversarial networks (GAN), variational autoencoders (VAE), and diffusion  are used in histopathology for various tasks, such as data augmentation~\cite{10.1007/978-3-031-58171-7_6}, anomaly detection~\cite{uzunova2019unsupervised}, and synthetic image generation~\cite{ yellapragada2023pathldmtextconditionedlatent}, for improving diagnostic models by addressing scarcity of data. GANs and diffusion generate realistic images to train models for tasks, such as cancer detection~\cite{LI2022102251}, while VAEs aid in anomaly detection by reconstructing healthy tissues and identifying deviations~\cite{guleria2023enhancing}. These models also facilitate style transfer between staining protocols, thus reducing costs and enhancing cross-modal comparisons. Overall, generative models enhance data efficiency and diagnostic accuracy in histopathology.

Evaluating the quality and performance of generative models is challenging due to the lack of well-established, theory-backed metrics. This complexity stems from the need to assess various aspects of generated images, such as quality, aesthetics, realism, and diversity, all of which are inherently subjective. While human evaluation is effective, it is costly, time-consuming, and impractical for large datasets. As a result, researchers rely on automated evaluation methods that should ideally respond consistently to image degradation, accurately measure image quality, and remain computationally efficient, particularly in resource-limited environments.

Commonly used metrics for evaluating generative models include Inception Score (IS)~\cite{NIPS2016_8a3363ab}, Kernel Inception Distance (KID)~\cite{bińkowski2018demystifying}, Fréchet Inception Distance (FID)~\cite{heusel2018ganstrainedtimescaleupdate}, and contrastive language-image pre-training (CLIP) Maximum Mean Discrepancy (CMMD)~\cite{jayasumana2024rethinkingfidbetterevaluation}. KID, FID and CMMD require thousands of real and generated images to compute a stable metric which becomes a big issue in medical domain with data scarcity. Flow-based Likelihood Distance~\cite{jeevan2024normalizingflowbasedmetricimage} have been proposed as an alternative, functioning effectively with fewer data, but they come with the complexity of training large normalizing flows, which is inefficient due to the need to maintain the large latent space. Another limitation of these metrics is their reliance on Inception-V3 features, which are trained on natural image datasets such as ImageNet. While effective for evaluating generative models that produce natural images, this approach performs poorly in domains, such as medical imaging, particularly in histopathology, where the image characteristics differ significantly from natural images.

Additionally, a quality metric trained on real patches of high quality can also be used to filter out low-quality patches from a whole slide image while training and testing deep learning pipelines for weakly supervised learning in histopathology~\cite{patil2023efficient}.







\section{ResNet-L2}

Normalizing flows are the only class of generative models capable of providing exact and efficient likelihood estimates for data~\cite{prince2023understanding}. However, they require the latent space to have the same dimensionality as the input data, leading to substantial computational demands, particularly for high-dimensional inputs, such as images. To address this issue, we apply normalizing flows to the extracted image features rather than the raw images, thus effectively reducing the dimensionality and computational complexity while preserving the model's ability to estimate data likelihoods accurately and efficiently.

Our proposed ResNet-L2 (RL2) metric employs a normalizing flow applied to image features extracted from a pre-trained ResNet-18, and computes the root mean square distance between the latent vectors of real and generated images. Specifically, we start with a set of real images, denoted as \(\mathcal{R}\), and a set of generated images, \(\mathcal{G}\) as shown in Figure~\ref{fig:rl2}. We use a pre-trained ResNet-18, truncated before the final layer, as the feature extractor \(f\), and a normalizing flow \(N\), parameterized by two trainable sets of parameters, \(\theta\) and \(\xi\) respectively.

\begin{equation}\label{eq:1}
    \textbf{x}_f = f(\textbf{x}_{in}, \theta) \hspace{5mm}\textbf{x}_{in}\in \mathcal{R}, \textbf{x}_f \in \mathbb{R}^{512}
\end{equation}
\begin{equation}\label{eq:2}
    \textbf{z} = N(\textbf{x}_f, \xi) \hspace{6mm}\textbf{z}\in \mathbb{R}^{512}, \textbf{x}_f \in \mathbb{R}^{512}
\end{equation}

We train the combined network  (Eq.~\ref{eq:1} and Eq.~\ref{eq:2}) on real images \(\mathcal{R}\) to maximize the log-likelihood of the real data. This training process is performed only once, allowing the network to map image features into a Gaussian latent space that assigns high likelihoods to real image features. Notably, since ResNet-18 is a significantly lighter model (11 M parameters) compared to Inception-V3 (23 M parameters) or CLIP, which are used in previous metrics, our approach offers faster and more computationally efficient metric computation. Moreover, by applying the normalizing flow to the extracted image features rather than the raw images, we further enhance efficiency due to the reduced dimensionality of the latent space, enabling quicker training. Training on real histopathology data allows the model to capture domain-specific features and assign higher probabilities to real histopathology images. This domain-specific adaptation is absent in previous metrics, which rely solely on frozen networks pre-trained on ImageNet.

After training the network on real images $\mathcal{R}$, we extract the mean of the latent vectors $\textbf{z}$ for real and generated images and compute the L2 (Euclidean distance) between the real and generated latent vectors\footnote{Code available at \url{https://github.com/pranavphoenix/RL2}}.

\begin{equation}
        \mathbf{z}_r = \dfrac{\sum_{z \in \mathcal{R}}  \mathbf{z}}{|\mathcal{R}|} \hspace{5mm}
        \mathbf{z}_g = \dfrac{\sum_{z \in \mathcal{G}}  \mathbf{z}}{|\mathcal{G}|} \hspace{5mm } \mathbf{z} \in \mathbb{R}^{512}
\end{equation}

\begin{equation}
    \text{RL2} = \|\mathbf{z}_r - \mathbf{z}_g\|_2 = \sqrt{\sum_{i=1}^{n} (z_{ri} - z_{gi})^2}
\end{equation}

The computation of L2 distance between two vectors is computationally lighter and faster than the computing Fréchet distance or maximum mean discrepancy between them. Hence, RL2 is lighter, faster and efficient to compute than previous metrics. 

\begin{figure}[htb]
\begin{minipage}[b]{1.0\linewidth}
  \centering
  \centerline{\includegraphics[scale=0.5]{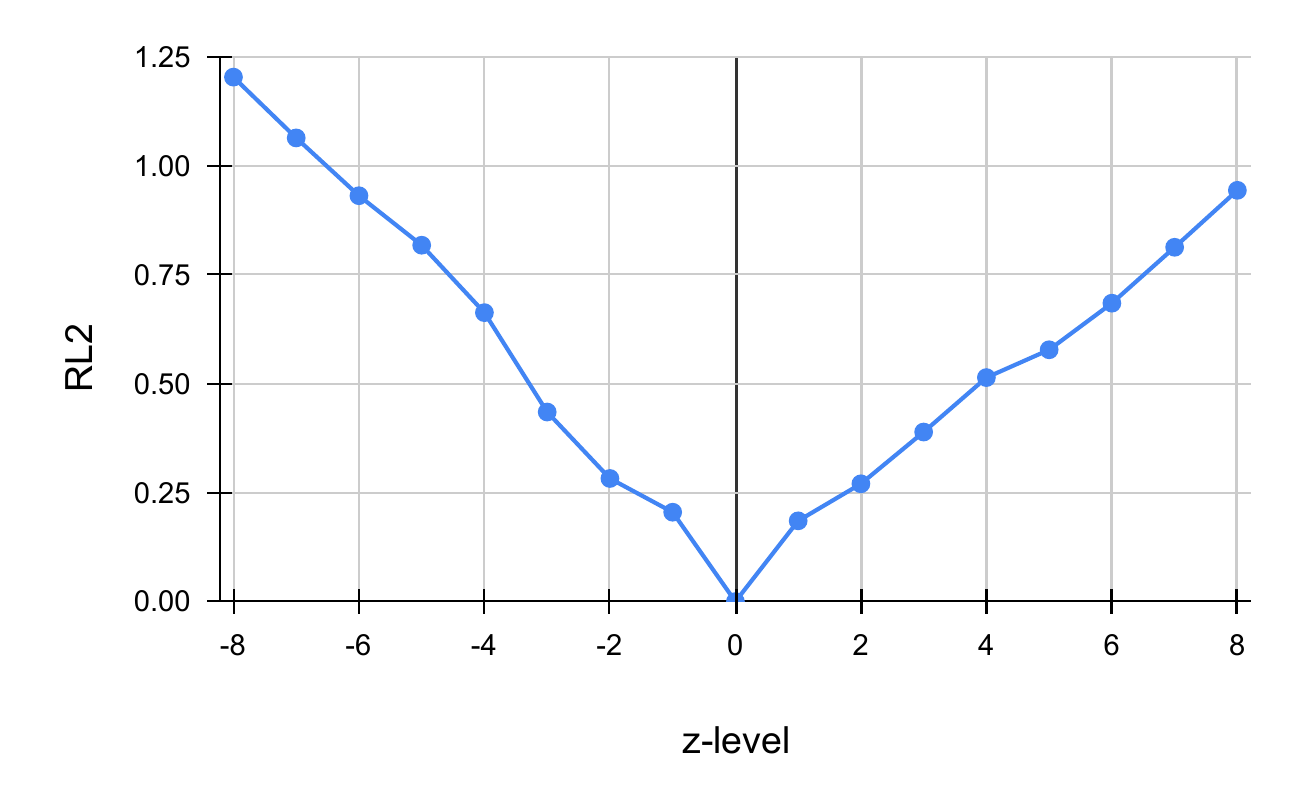}}
\end{minipage}
\begin{minipage}[b]{1.0\linewidth}
  \centering
  \centerline{\includegraphics[scale=0.3]{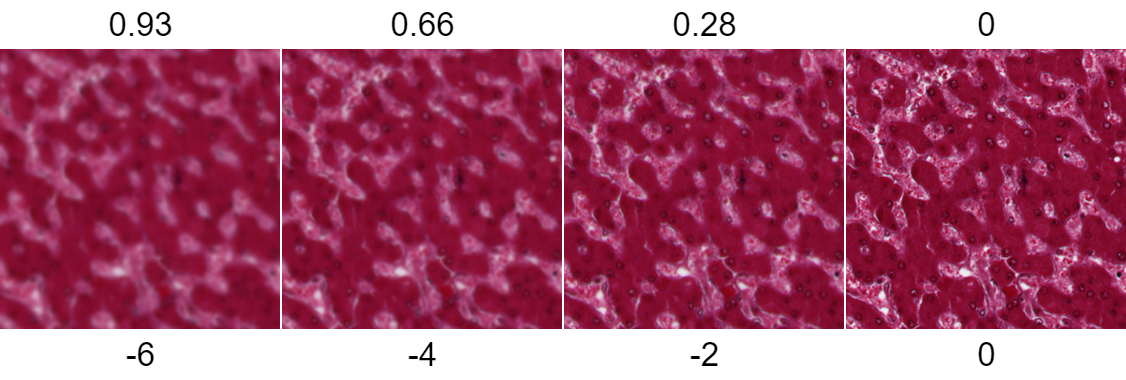}}
\end{minipage}
\caption{The behaviour of RL2 shows that it is monotonic with increasing levels of blur caused by variation of z-levels in FocusPath dataset. The top row shows the RL2 value and the bottom row shows the z-values.}
\label{fig:zlevel}
\end{figure}

\section{Datasets, Implementation and Experiments}

\begin{figure}[]
\begin{minipage}[b]{1.0\linewidth}
  \centering
   \centerline{\includegraphics[scale=0.5]{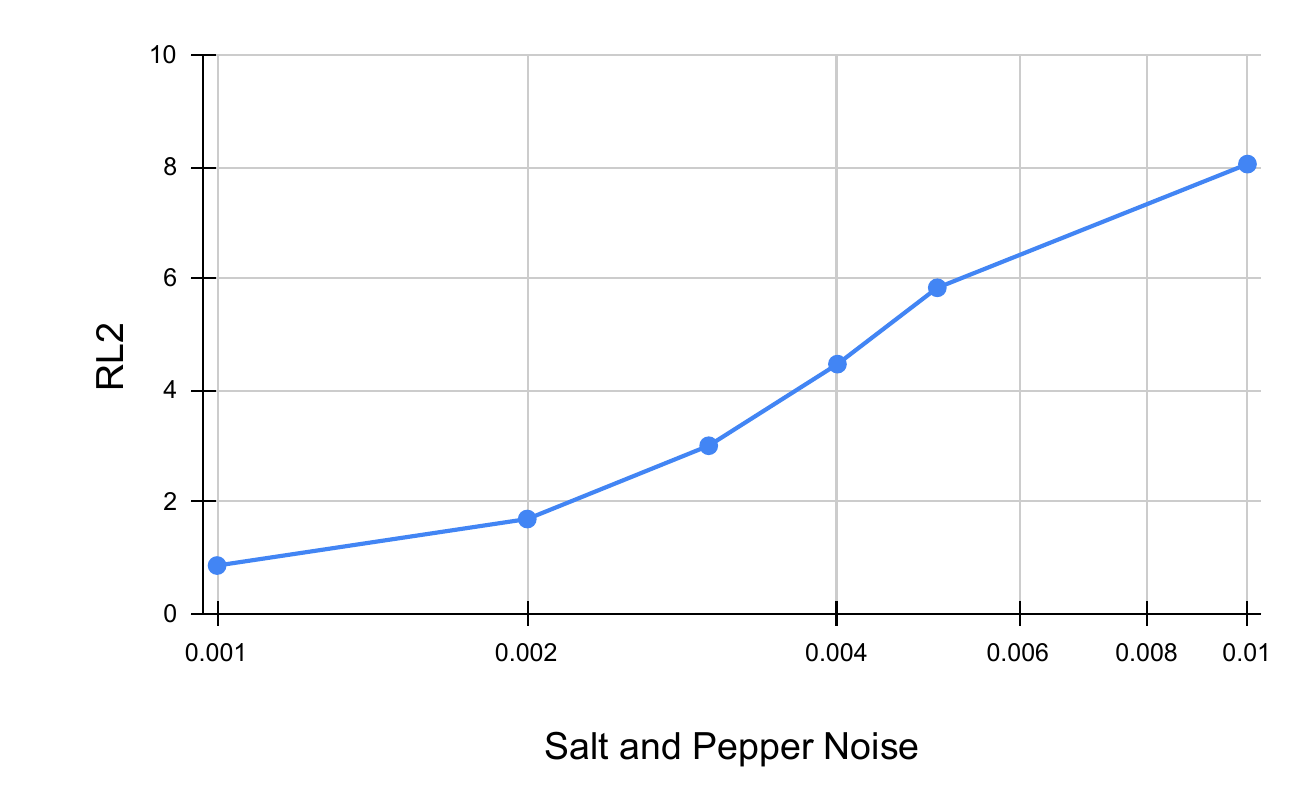}}
\end{minipage}
\begin{minipage}[b]{1.0\linewidth}
  \centering
  \centerline{\includegraphics[scale=0.25]{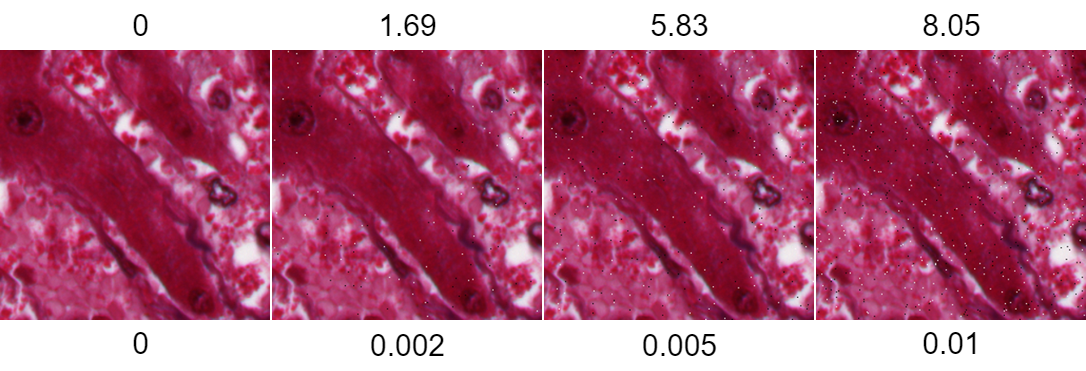}}
\end{minipage}
\caption{The behaviour of RL2 shows that it is monotonic with increasing levels of salt and pepper noise in histopathology images from $0$ z-level FocusPath dataset.. The top row shows the RL2 value and the bottom row shows the noise values.}
\label{fig:salt}

\end{figure}



\begin{figure}[]
\begin{minipage}[b]{1.0\linewidth}
  \centering
   \centerline{\includegraphics[scale=0.5]{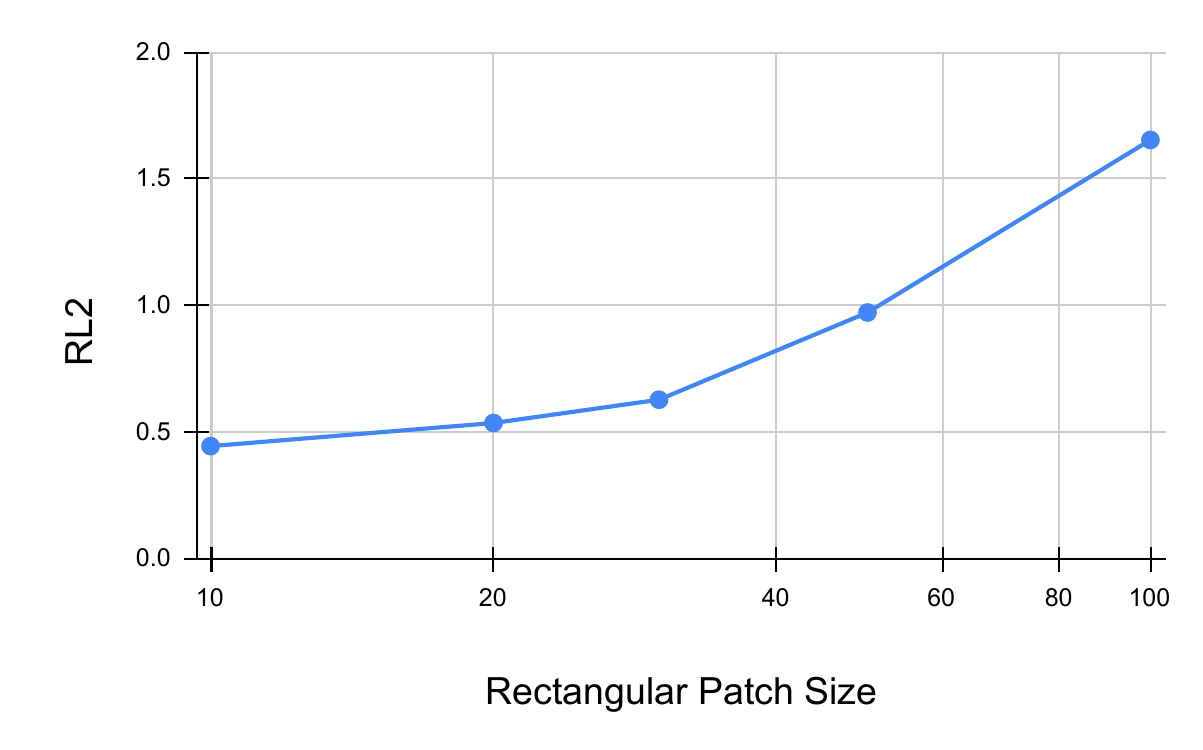}}
\end{minipage}
\begin{minipage}[b]{1.0\linewidth}
  \centering
  \centerline{\includegraphics[scale=0.25]{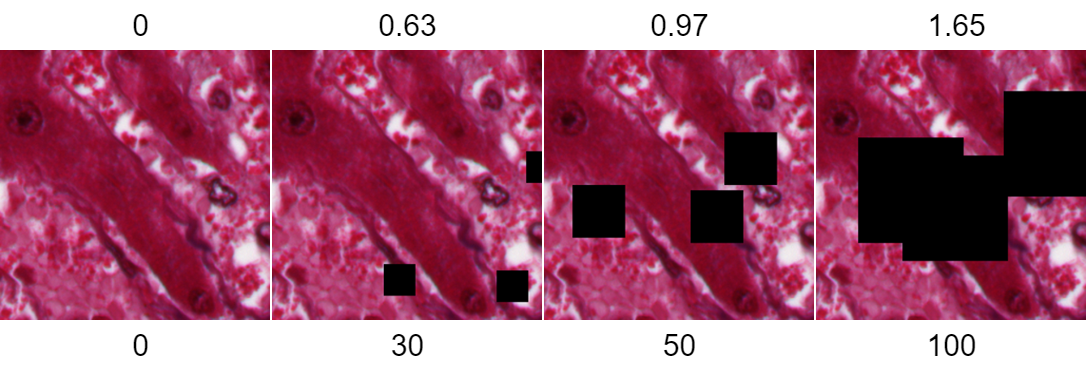}}
\end{minipage}
\caption{The behaviour of RL2 shows that it is monotonic with increasing levels of rectangular patch noise in histopathology images from $0$ z-level FocusPath dataset.. The top row shows the RL2 value and the bottom row shows the noise values.}
\label{fig:patch}

\end{figure}

\begin{figure}[htb]
  \centering
   \centerline{\includegraphics[scale=0.5]{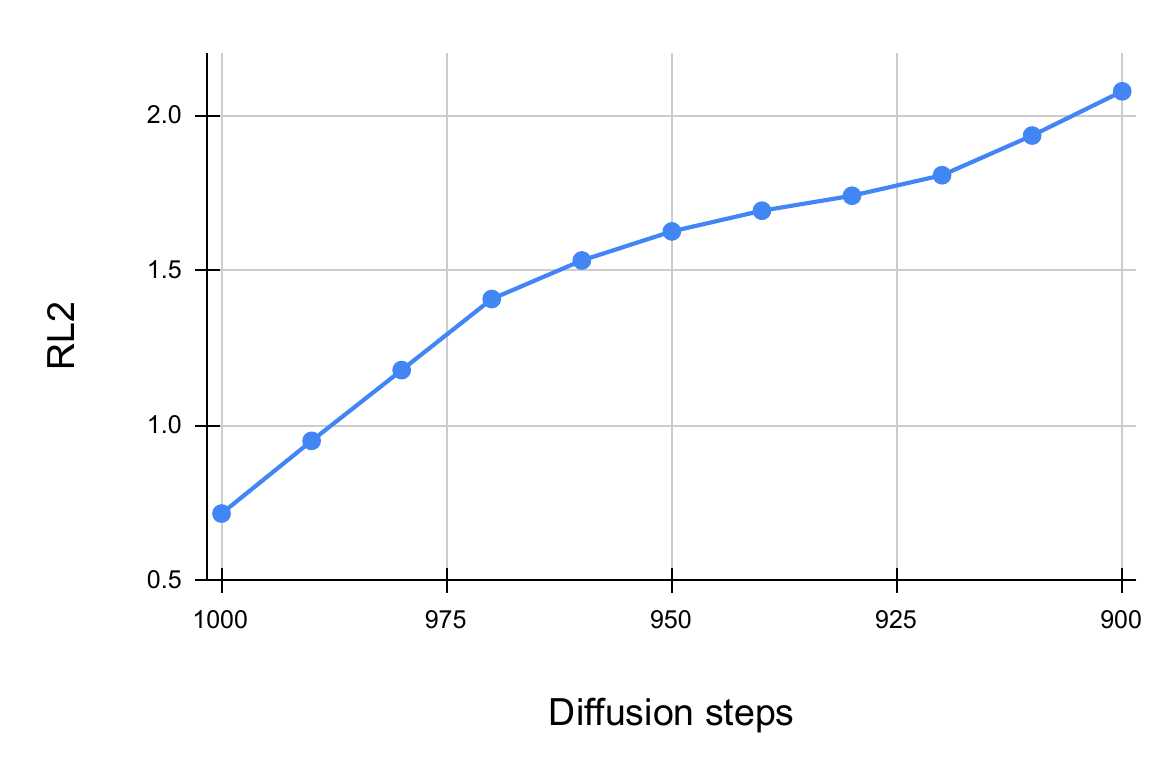}}
\caption{The behaviour of RL2 shows that it is monotonic with increasing levels of salt and pepper noise in histopathology images from $0$ z-level FocusPath dataset.}
\label{fig:diffusion}

\end{figure}

The FocusPath dataset~\cite{8672094} consists of 864 image patches, each with a resolution of $1024\times1024$ pixels in sRGB format, capturing varying degrees of focus. These patches are cropped from nine distinct whole slide images (WSIs), with 16 different z-levels employed to simulate various out-of-focus conditions. The tissue slides selected for the dataset feature diverse color staining techniques, ensuring that the dataset reflects a wide range of histopathological staining variations. The dataset involves image patches with varying degrees of focus, translating to different levels of blur. This dataset serves as an excellent tool for assessing our metric's capability to distinguish between blurry and clear images. Moreover, it allows us to evaluate the monotonicity of our metric with respect to the degree of blur, as it should assign higher values to images with greater blurriness. We train our network only on images with z-level $0$ (which are blur free) and evalaute it on all the other images with higher degree of blur. We add salt and pepper noise and rectangular patch noise to the $0$ z-level images for our experiments involving other kinds of noise. We used images of $100\times$ resolution from BreakHis~\cite{7312934} dataset for generating synthetic images for our experiments to test RL2 for diffusion noise. 

For the experiment to filter out low-quality noisy image patches from good quality ones, we used HistoROI dataset~\cite{patil2023efficient}. HistoROI dataset is developed to segment WSIs into six key classes: epithelium, stroma, lymphocytes, adipose, artifacts, and miscellaneous. Artifacts in this dataset include out-of-focus areas, tissue folds, cover slips, air bubbles, pen marks, and areas with extreme over- or under-staining, which were carefully labeled to aid in quality control for pathology image analysis. We trained our network using 1500 artifact-free patches from epithelium and stroma classes. We then use this network to classify 474 artifact patches from 474 clean patches. 

We use a normalizing flow using real-valued non-volume preserving (real NVP) transformations~\cite{dinh2017densityestimationusingreal} in our experiments. We utilized $256\times256$ patches from all the datasets, achieved by cropping the original images. All experiments were run on Nvidia A6000 GPU. For experiments using salt and pepper noise, rectangular patch noise, and diffusion noise, we used the same experimental setup used by FID paper~\cite{heusel2018ganstrainedtimescaleupdate}.

\section{Results and Discussion}

Figure~\ref{fig:zlevel} illustrates that as the level of blur increases, indicated by higher z-levels, ranging from $1$ to $8$ and $-1$ to $-8$, our metric RL2 effectively identifies these images as noisier compared to the $0$ z-level images. Moreover, RL2 displays a monotonous increase with greater intensity of blur in both directions, highlighting its effectiveness as an ideal tool for identifying blurry images compared to clear, noise-free ones.

A robust evaluation metric for generative models must effectively distinguish between varying levels of different noise types. In our experiments with salt-and-pepper noise and rectangular patch noises, common in histopathology images, our metric, RL2, shows a monotonic increase with rising noise levels as shown in Figure~\ref{fig:salt} and Figure~\ref{fig:patch}. This demonstrates its effectiveness in evaluating image quality. The capability to identify corruptions in real images makes our metric a valuable tool for detecting subtle differences caused by various noises. Even when synthetic image data significantly differs from real data, our metric reliably identifies varying levels of various noise.

The results of our evaluation metric on the diffusion process, are illustrated in Figure~\ref{fig:diffusion}. As shown, an increase in the number of diffusion steps correlates with a reduction in noise levels, a pattern our metric effectively captures through lower values. This monotonic reduction highlights quality improvements in the diffusion process, essential for accurate evaluation of histopathological images from generative models.

In the experiment for filtering out low-quality patches with artifacts from high quality patches, our network trained only on clean patches gives an AUC score of 0.76. Even without being trained on the artifacts, the model is able to identify clean patches from noisy ones based on the their likelihood.

RL2 metric achieves stability with just 300 samples compared to traditional methods which FLD required over 20,000 images to reach a stable value~\cite{jayasumana2024rethinkingfidbetterevaluation}. This significant sample efficiency of RL2 enables it to produce reliable, stable metric values with minimal data.

\section{Conclusions}

Popular generative model evaluation metrics, such as FID, have proven unreliable for assessing the latest models, such as diffusion. This is particularly true in domains, such as histopathology, where the available images are limited. Previous metrics requiring thousands of samples for stability is not suitable, nor are those metrics relying on pre-trained ImageNet weights due to the distinct nature of histopathology image features. 

We developed a new evaluation metric based on normalizing flow and measures the L2 distances between real and generated features. Our metric has been shown to be monotonic with respect to various noise types found in histopathology, including noise, occlusion, and blur. Additionally, because our metric is lighter, faster and requires significantly fewer resources, it is efficient than previous metrics. We believe this work will empower researchers to train and test generative models more efficiently.

\section{Compliance with Ethical Standards}
This research used public data, ethical approval was not required as confirmed by the license attached with the data.

\bibliography{references}
\bibliographystyle{unsrt}


\end{document}